\documentclass[12pt]{article}
\usepackage{a4wide}
\usepackage{amssymb}
\usepackage{graphicx}
\begin{document}
{\renewcommand{\thefootnote}{\fnsymbol{footnote}}
\begin{center}
{\LARGE   Black-hole models in loop quantum gravity}\\
\vspace{1.5em}
Martin Bojowald\footnote{e-mail address: {\tt bojowald@gravity.psu.edu}}
\\
\vspace{0.5em}
Institute for Gravitation and the Cosmos,\\
The Pennsylvania State
University,\\
104 Davey Lab, University Park, PA 16802, USA\\
\vspace{1.5em}
\end{center}
}

\setcounter{footnote}{0}

\begin{abstract}
Dynamical black-hole scenarios have been developed in loop quantum
  gravity in various ways, combining results from mini and midisuperspace
  models. In the past, the underlying geometry of space-time has often been
  expressed in terms of line elements with metric components that differ from
  the classical solutions of general relativity, motivated by modified
  equations of motion and constraints. However, recent results have shown by
  explicit calculations that most of these constructions violate general
  covariance and slicing independence. The proposed line elements and
  black-hole models are therefore ruled out. The only known possibility to
  escape this sentence is to derive not only modified metric components but
  also a new space-time structure which is covariant in a generalized
  sense. Formally, such a derivation is made available by an analysis of the
  constraints of canonical gravity, which generate deformations of
  hypersurfaces in space-time, or generalized versions if the constraints are
  consistently modified. A generic consequence of consistent modifications in
  effective theories suggested by loop quantum gravity is signature change at
  high density. Signature change is an important ingredient in long-term
  models of black holes that aim to determine what might happen after a black
  hole has evaporated. Because this effect changes the causal structure of
  space-time, it has crucial implications for black-hole models that have been
  missed in several older constructions, for instance in models based on
  bouncing black-hole interiors. Such models are ruled out by signature change
  even if their underlying space-times are made consistent using generalized
  covariance. The causal nature of signature change brings in a new internal
  consistency condition, given by the requirement of deterministic behavior at
  low curvature. Even a causally disconnected interior transition, opening
  back up into the former exterior as some kind of astrophysical white hole,
  is then ruled out. New versions consistent with both generalized covariance
  and low-curvature determinism are introduced here, showing a remarkable
  similarity with models developed in other approaches, such as the
  final-state proposal or the no-transition principle obtained from the
  gauge-gravity correspondence.
\end{abstract}

\newpage

\section{Introduction}

Black-hole models have recently regained considerable attention in loop
quantum gravity. A certain consensus seems to have formed according to which
the singularity in a classical black hole is replaced by a non-singular phase
in which the density and curvature are not infinite but large, such that
infalling matter might bounce back and re-emerge after the initial horizon has
evaporated. Such models are not only used in conceptual discussions about
a possible non-singular fate of black holes, but even in phenomenological
descriptions that aim to derive potentially observable effects from the
re-emergence of matter.

It is important to note that none of these models are based on consistent
embeddings of possible effects from loop quantum gravity in a covariant
space-time theory. Rather, these models assume that effects suggested by some
equations of loop quantum cosmology, such as bounded density or curvature, can
be modeled reliably by modified line elements that amend the singular
solutions of general relativity in various forms. Therefore, these models
implicitly assume, but do not show, that loop quantum gravity has a
well-defined semiclassical description in which its dynamics can be described
by space-time equipped with Riemannian geometry. Since loop quantum gravity
aspires to be a background-independent approach to quantum gravity, however,
the structure of space-time cannot be presupposed but should rather be derived
from the theory. Current models of black holes therefore have important
conceptual lacunae.

Provided models are sufficiently controlled for a space-time analysis to be
possible, which in practice means that they have a sufficiently developed
canonical structure, general covariance can be tested by various means.
Recently, it has been shown by direct calculations that all current
loop-inspired black-hole models of bounce form violate general covariance and
therefore fail to describe space-time effects in any meaningful
way 
\cite{SphSymmCov,TransComm,NonCovDressed,Disfig}.  It is important to note
that these results were obtained directly within the proposed models. They are
therefore independent of any difference between approaches that have been put
forward in order to formulate inhomogeneous models of loop quantum gravity,
such as hybrid models \cite{Hybrid}, the dressed-metric approach \cite{AAN},
partial Abelianization in spherically symmetric models \cite{LoopSchwarz}, or
using timelike homogeneous slices in static spherically symmetric space-times
\cite{Transfig}, just to name those to which the recent no-go results about
covariance directly apply.

Fortunately, even before these results became available, it had been
found that covariance can be preserved by some loop effects at least in a
deformed way, which respects the number of classical symmetries underlying
general covariance but may change their algebraic relationships
\cite{ConstraintAlgebra,JR,LTBII,HigherSpatial,SphSymmOp,ScalarHol,ScalarHolInv,Action,DeformedCosmo}.
Deformation of the algebra, as opposed to violation of some gauge
transformations as found in most of the approaches mentioned in the preceding
paragraph, makes sure that the theory remains background independent in the
sense that transformations remove the same number of gauge degrees of freedom
as in the classical theory. However, as a consequence, space-time is rendered
non-Riemannian --- unless field redefinitions are applied in certain cases ---
and at present no universal non-Riemannian space-time geometry has been
identified that could describe all the deformation effects. 

Such a deformed theory is still predictive in principle because it has a
consistent canonical formulation that defines unambiguous observables. But the
relationship between gauge invariance and slicing independence (according to a
suitable space-time structure) is less obvious than in the classical limit,
complicating interpretations of space-time effects such as black holes.
The resulting non-Riemannian structure would have to deviate from standard
constructions in very basic ways because it is modified even in properties of
generic tensorial objects encoded by the tensor-transformation law (which is
replaced by gauge transformations in a canonical formulation). Therefore,
deformed geometrical properties cannot be captured completely by well-known
classical ingredients such as torsion or non-metricity.

Possible space-time structures consistent with deformed symmetries are still
being explored. For instance, it is known that deformations of the form
implied by effects from loop quantum gravity are different from modified
coordinate transformations found in non-commutative \cite{NCHDA} or
multifractional geometry \cite{FractHypDef}. There are some relationships with
deformed Poincar\'e symmetries in suitable limits in which a Minkowski
background is obtained \cite{DeformedRel,DeformNonComm}, which confirms that
Lorentz transformations are not necessarily violated, but the Minkowski
reduction of general space-time transformations is rather strong and cannot be
sufficient for a complete understanding of deformed space-time structures.
As we will show here, what is known about the resulting consistent space-time
structures suggests a markedly different scenario of non-singular black holes,
compared with bounce-based black holes.

We will first review current proposals in Section~\ref{s:Proposals} and point
out their hidden assumptions and other weaknesses. (Since our focus will be on
dynamical aspects and space-time structure, we will not discuss results about
black-hole entropy in loop quantum gravity. See \cite{PerezBlackHoles} for a
recent review of this topic.) Section~\ref{s:Mod} will then introduce basic
aspects of modified space-time structures in spherically symmetric models.
General covariance is analyzed in detail in the two main canonical
loop-modified approaches, \cite{LoopSchwarz} and \cite{Transfig}, in which
modified line elements have been proposed, and found to be lacking. But the
former proposal unwittingly hints at an important role played by deformed
space-time structures and signature change \cite{SphSymmCov}, found
independently (and earlier) by direct studies of consistent constraints in
spherically symmetric models \cite{JR,LTBII,HigherSpatial,SphSymmOp}.

Signature change is a generic implication of modified space-time structures in
models of loop quantum gravity, which can replace classical singularities but
may lead to other unwanted implications such as indeterministic behavior even
at low curvature. (The role of signature change and possible evasions will
also be discussed.) The condition that physics at low curvature be
deterministic then rules out certain black-hole models, including bouncing
ones, and suggests new ones that are compatible with determinism as well as
generalized covariance. These scenarios show interesting relationships with
other proposals unrelated to loop quantum gravity. There is therefore a
refreshing contrast with bounce-based black holes in loop quantum gravity,
which are often put in opposition to other approaches.

\section{Proposals}
\label{s:Proposals}

Recent interest in a certain type of black-hole models in loop quantum gravity
was rekindled by the discussion in \cite{PlanckStar} which suggested that ``a
strong short-scale repulsive force due to quantum effects,'' motivated by
certain claims in loop quantum cosmology, might cause collapsing matter in a
black hole to bounce back well before evaporation could have reduced the black
hole mass to a tiny Planckian value. Here, we will not be concerned with the
question of whether related phenomenological studies are meaningful. We will
rather address the more basic conceptual question of whether the proposal, and
in particular its claimed relationship with loop quantum cosmology, can be
realized in a consistent space-time framework that respects general
covariance. (Our discussion will also be independent of the dubious claim that
loop quantum effects imply a ``strong short-scale repulsive force;'' see
\cite{Claims}.)

\subsection{Basic premise of bounce-based black holes}
\label{s:Basic}

Much of the analysis in \cite{PlanckStar} is based on a postulated line
element 
\begin{equation}
 {\rm d}s^2= -F(r){\rm d}u^2+ 2{\rm d}v{\rm d}r+ r^2({\rm
   d}\vartheta^2+\sin^2\vartheta{\rm d}\varphi^2)
\end{equation}
that modifies the classical Schwarzschild solution by the inclusion of an
additional term in the function
\begin{equation} \label{Fr}
 F(r)=1-\frac{2mr^2}{r^3+2\alpha^2m}
\end{equation}
if $\alpha\not=0$. Also here, we will not be interested in the question of
whether this functional form is justified based on models of loop quantum
gravity. A more basic question is whether {\em any} modification suggested by
models of loop quantum gravity can be compatible with general covariance, such
that its dynamical solutions are consistent and can be expressed in terms of a
well-defined space-time line element.  The specific function (\ref{Fr}) was
first proposed in \cite{ClosedHorTrapping}, where it was also shown that it
may be obtained as a solution of a covariant theory, such as general
relativity with a suitable stress-energy tensor that falls off quickly with
increasing $r$. The question we pose here is more fundamental and works in the
opposite direction: Starting with a specific modified theory or at least a set
of model equations that are not gauge-fixed so as to keep space-time
properties accessible, is it possible to express its solutions in the form of
Riemannian line elements consistent with gauge transformations.

This question is highly non-trivial because modifications in models of loop
quantum cosmology are first implemented in the Hamiltonian constraint of the
theory. This constraint generates gauge transformations which, together with
the transformations generated by the diffeomorphism constraint, are equivalent
to space-time coordinate changes in the classical theory. If the Hamiltonian
constraint is modified, gauge transformations may be non-classical, and it is
no longer guaranteed that, when applied to metric components, they remain dual
to coordinate changes of coordinate differentials ${\rm d}x^a$. If this is not
the case, the expression ${\rm d}s^2=g_{ab}{\rm d}x^a{\rm d}x^b$ is not
invariant, and fails to define a meaningful line element on which geometry
could be based. We conclude that two non-trivial properties must be shown for
any desired space-time effect, such as a bouncing black-hole interior, to be
meaningful:
\begin{enumerate}
\item[(i)] It must be possible to obtain the effect as a specific solution
  $g_{ab}$ of a consistent set of field equations.
\item[(ii)] Together with the specific solution required by (i), there must be
  a set of solutions $g_{a'b'}$ related to $g_{ab}$ by gauge transformations
  that 
\begin{enumerate}
 \item[(a)] preserve the field equations and 
\item[(b)] have corresponding coordinate transformations
  from $x^a$ to $x^{a'}$ such that $g_{a'b'}=(\partial x^a/\partial
  x^{a'})(\partial x^b/\partial x^{b'}) g_{ab}$.
\end{enumerate}
\end{enumerate}

Each of these three conditions --- (i), (ii.a) and (ii.b) --- is non-trivial
and must be checked carefully for any proposed modification of general
relativity that is not of higher-curvature, scalar-tensor or some related
form. The set of solutions $g_{a'b'}$ in (ii) should be sufficiently large to
include all geometries of the desired form, such as spherically symmetric
metrics in models of non-rotating black holes.  Condition (ii.b) then ensures
that an invariant line element ${\rm d}s^2=g_{ab}{\rm d}x^a{\rm d}x^b$ can be
constructed from solutions of the field equations. This condition is therefore
crucial, but it has often been overlooked. There are models proposed in loop
quantum gravity in which line elements are used even though none of the
conditions (i), (ii.a) and (ii.b) have been checked. Two such examples
\cite{PlanckStar,Fireworks} are briefly discussed below. While other models
propose at least some form of field equations in line with condition (i), the
two remaining conditions, (ii.a) and (ii.b), have rarely been checked
explicitly. For instance, \cite{Transfig}, discussed in
Section~\ref{s:Slicing} below, checked neither (ii.a) nor (ii.b), but it is
now known that it is impossible for both conditions to be realized in such
constructions \cite{TransComm,Disfig}. The model proposed in
\cite{LoopSchwarz}, discussed in Section~\ref{s:SphSymm} below, successfully
checked conditions (i) and (ii.a), but left (ii.b) open. As shown in
\cite{SphSymmCov}, condition (ii.b) is, in fact, not met by the model of
\cite{LoopSchwarz}, but a weakened form, unnoticed in \cite{LoopSchwarz}, can
be derived in which the classical structure of space-time, implemented in
(ii.b) by reference to the tensor-transformation law of Riemannian geometry,
is modified. This result is an example of modified space-time structures
discussed in detail in Section~\ref{s:Mod}. At present, no model is known in
loop quantum gravity that obeys all three conditions (i), (ii.a) and (ii.b)
without a generalization from (ii.b) to
\begin{enumerate}
\item[(ii.b')] The gauge transformations are such that their classical limit
  has corresponding coordinate transformations from $x^a$ to $x^{a'}$ with
  $g_{a'b'}=(\partial x^a/\partial x^{a'})(\partial x^b/\partial x^{b'})
  g_{ab}$.
\end{enumerate}
Even though the classical theory is known to be covariant and consistent with
the tensor-transformation law, experience with models of loop quantum gravity
shows that embedding the condition (ii.b') in a theory with modified gauge
transformations is non-trivial because condition (ii.a) must be fulfilled
before the classical limit can be taken in order for the modified theory to be
well-defined.

In \cite{PlanckStar}, which violates all three conditions, it is taken for
granted that quantum-gravity effects can always be described by modified line
elements. For instance, the proposal of an astrophysical object of Planckian
density is first described as ``The main hypothesis here is that a star so
compressed would not satisfy the classical Einstein equations anymore, even if
huge compared to the Planck scale.'' After further specifications, the paper
continues with ``Let us write a metric that could describe the resulting
effective geometry.'' Here, the justified assumption that Einstein's equation
may be modified at high density because of quantum-gravity effects is directly
turned into the unsupported postulate that the corresponding geometry should
be Riemannian, described by a metric tensor that determines coefficients in a
line element. Similarly, \cite{Fireworks}, which presents a more refined
metric for bounce-based black holes, erroneously states that ``the technical
result of the present paper is that such a metric exists for a bouncing black
to white hole'' even though the paper did not actually show that a metric of
any kind exists that could be used to describe the desired effect as a
solution of covariant equations.  The same paper concludes with statements
such as ``the metric we have presented poses the problem neatly for a quantum
gravity calculation. The problem now can be restricted to the calculation of a
quantum transition in a finite portion of spacetime'' and ``this is precisely
the form of the problem that is adapted for a calculation in a theory like
covariant loop quantum gravity'', claiming that ``the spinfoam formalism is
designed for this.'' The calculation of a quantum transition amplitude is not
sufficient because it must first be shown that a quantum theory of gravity
used in such a derivation does, in fact, allow a metric structure to describe
its solutions. This task has not been performed in the spin-foam formalism or
loop quantum gravity, in spite of the epithet ``covariant'' assigned to it in
the preceding quote.

The tacit assumption that any solution of quantum gravity must be of metric
form fails to recognize the non-trivial nature of the availability of line
elements. In particular in background-independent approaches to quantum
gravity, the structure of space-time is to be derived, not to be assumed. A
detailed analysis should then be performed to see whether line elements are
available. This conclusion refers to line elements of a generic form, setting
aside the question of what their precise coefficients might be. 

Spin-foam models are ill-suited for questions about space-time structure
because it has not even been shown whether they are consistent space-time
theories. In particular, it has not been shown that their discrete
path-integral measure is covariant; see \cite{Anomaly}. The canonical
formulation of (quantum) gravity is better equipped to analyze covariance
questions because it is closely related to the general consistency condition
that constraints or their quantizations should be first-class and free of
anomalies. Checking this condition may be complicated, but it is well-defined
and amenable to systematic methods. The canonical formulation also allows one
to work out effective descriptions which take into account detailed properties
of quantum states \cite{EffAc,Karpacz,EffCons}. We will first review salient
features of such effective equations, and then return to the question of
covariance.

\subsection{Modifications suggested by loop quantum cosmology}

In models of loop quantum gravity, modifications of the classical equations
for black holes are usually based on what has been studied for some time in
isotropic cosmological systems. In this context, the Friedmann equation
\begin{equation} \label{Friedmann}
 \left(\frac{\dot{a}}{a}\right)^2=\frac{8\pi G}{3} \rho
\end{equation}
for the scale factor $a$ is first presented in canonical form,
\begin{equation} 
 C=V\rho- 6\pi G V p_V^2=0
\end{equation}
where
\begin{equation} 
 p_V= -\frac{1}{4\pi G} \frac{\dot{a}}{a}
\end{equation}
is canonically conjugate to the volume, $V=a^3$.  The constraint $C=0$
therefore replaces the Friedmann equation. 

In loop quantum cosmology \cite{ROPP,Foundations}, it is argued that periodic
functions of $p_V$ (or of some other combination of the canonical variables,
usually linear in $p_V$ but not necessarily in $V$) should be used instead of
polynomials, modeling matrix elements of holonomies for compact groups (in
particular, the spatial rotation group used in loop quantum gravity). A
modified constraint of the form
\begin{equation} \label{Cmod}
 C_{\rm modified}= V\rho-6\pi GV \frac{\sin^2(\delta p_V)}{\delta^2}
\end{equation}
with an ambiguity parameter $\delta$ then implies that the energy density is
bounded on solutions of $C_{\rm modified}=0$. If this constraint is used
without any further quantum corrections, Hamilton's equations generated by
$C_{\rm modified}$ can be rewritten as a modified Friedmann equation
\cite{AmbigConstr}
\begin{equation}  \label{ModFried}
  \left(\frac{\dot{a}}{a}\right)^2= 
  \frac{8\pi G}{3} \left(\rho-\frac{\rho^2}{\rho_{\rm max}}\right)
\end{equation}
with the maximum density 
\begin{equation}
 \rho_{\rm max}= \frac{6\pi G}{\delta^2}
\end{equation}
implied on solutions of the modified constraint $C_{\rm modified}=0$. If
$\rho=\rho_{\rm max}$, $\dot{a}=0$ and the scale factor has a turning point.

The modified Friedmann equation (\ref{ModFried}) indicates that bounces may be
possible in models of loop quantum gravity. However, it does not present
conclusive evidence because it incorporates only one out of several possible
quantum effects. It does represent the characteristic loop behavior of
isotropic models \cite{IsoCosmo} because the classical quadratic dependence of
the constraint on the momentum is replaced by a periodic function, which can
be interpreted as a matrix element of a gravitational holonomy. The latter,
rather than momentum components themselves, are represented as operators on
the kinematical Hilbert space of loop quantum gravity \cite{LoopRep,ALMMT}. An
analogous property is encoded in the modification given by $C_{\rm modified}$.

Because loop quantum cosmology is a {\em quantum} theory, however, one also
expects that general quantum effects, such as dynamical implications of
fluctuations and higher moments of a state, should be relevant, in particular
at high density where the simple (\ref{ModFried}) seems to imply a bounce
\cite{BouncePert,BounceSqueezed}. These quantum effects may be ignored in a
low-curvature universe model at late times, but they are highly relevant (and
suppressed in the simple (\ref{ModFried})) close to a spacelike singularity
\cite{Infrared,EFTLQC}.  A detailed analysis has also revealed previously
unrecognized ambiguities in a quantum version of $C_{\rm modified}$ because
the periodic function, combined with $V$, can be implemented by several
inequivalent representations of the Lie algebra ${\rm sl}(2,{\mathbb R})$
\cite{NonBouncing}. It therefore remains unclear how generic bounces in loop
quantum cosmology are, and by extension bounces of black-hole interiors; see
\cite{Claims} for a detailed analysis.

Another major problem is relevant in loop-based models of black holes. The
parameter $\delta$ in $C_{\rm modified}$, which is crucial in determining
properties of bouncing solutions, is related to the characteristic scale of a
specific kind of quantum effect. Its presence is motivated by the application
of holonomy operators in the full theory of loop quantum gravity, in which
case $\delta$ is related to the length of a curve along which parallel
transport is computed. In proposed constructions of Hamiltonian constraint
operators
\cite{QSDI,ThreeDeform,AnoFreeWeak,TwoPlusOneDef,TwoPlusOneDef2,OffShell,ConstraintsG},
the curve is related to links of a state acted on by the holonomy, expressed
in the spin-network basis. If there is any relationship between loop quantum
cosmology and loop quantum gravity, the single parameter $\delta$ therefore
has to encode detailed properties of an underlying dynamical state relevant
for cosmological or astrophysical evolution.

Needless to say, it is at present impossible to derive a value for $\delta$
from the full theory, but one may nevertheless use such a modification to
study possible outcomes of the loop representation. However, for reliable
conclusions the parameter should be put into model equations in a sufficiently
general form. In particular, the value of $\delta$ may have to be adjusted, or
renormalized, as the underlying space-time state evolves. Instead of a
constant $\delta$, one should therefore use a function that depends on a
relevant scale, such as the energy density or curvature in a certain range of
evolution. As the scale changes, $\delta$ does too, which may be modeled as a
certain function of $a$, a simpler parameter related to the energy scale on
solutions of the Friedmann equation. Again, it is at present impossible to
derive a specific function $\delta(a)$ from a space-time state, and therefore
a sufficiently general ansatz is required for generic conclusions.

The large freedom in choosing such a function, compared with a single
constant, implies that it is impossible to justify claims about detailed
long-term effects in loop models of bounce-based black holes.  For instance,
one of the claims made in \cite{Transfig} states that ``if the radius of the
black hole horizon in asymptotic region I [before the bounce] is, say, $r_{\rm
  B} = 3\: {\rm km}$, corresponding to a solar mass, that of the white hole
horizon in asymptotic region III [after the bounce] is $r_{\rm W} \approx (3 +
{\cal O}(10^{-25}))\: {\rm km}$.''  Unfortunately, this statement is derived
from equations that use analogs of $\delta$ which are constant on solutions of
the constraint and equations of motion and therefore ignore
renormalization. The proposal of \cite{Transfig} simply assumes that a single
effective theory with constant parameters can be used over a vast range of
scales, stretching from the low-curvature near-horizon region of a $3\:{\rm
  km}$ black hole all the way to Planckian curvature at the putative bounce,
and back to low curvature at a white-hole horizon of similar size. The
statement not only ignores the possibility (or necessity) of renormalization,
but also expects that maintaining a relative precision of $10^{-25}$ is
believable over such a vast range of scales, including quantum-gravity
regimes.

These problems make it difficult to draw justified conclusions about
bounce-based black holes in models of loop quantum gravity. However, such
models are still useful because they have revealed, somewhat unintentionally,
that the effects taken from loop quantum gravity require non-trivial
space-time structures in order to be meaningful. These are local space-time
structures, related to the form of general covariance realized in the
models. Because they are local, they are not sensitive to questions of
long-term quantum evolution or renormalization, and they can be parameterized
in sufficiently general form to include some quantization
ambiguities. (On occasion, they also rule out certain ambiguities by the
condition of covariance.) We will now turn to these fundamental questions,
continuing in this section with canonical models that have been proposed for
bounce-based black holes.

\subsection{Violations of general covariance}

As a canonical version of the conditions given in Section~\ref{s:Basic}, the
task is to find consistent versions of the constraints of general relativity,
modified such that they can incorporate bounce effects and at the same time
respect general covariance in the canonical form of hypersurface
deformations. We will now illustrate the highly non-trivial nature of the
combination of these conditions.

\subsubsection{Slicing dependence}
\label{s:Slicing}

Modifications similar to those in $C_{\rm modified}$ can be implemented in
anisotropic models \cite{HomCosmo,SchwarzN}, including Kantowski--Sachs
space-times which may describe the interior of Schwarzschild-type black holes
where the timelike Killing vector field of the static exterior is replaced by
a spacelike field within the horizon, implying homogeneity. On this basis, a
long-standing suggestion is that collapsing space-time in a black hole could
bounce back at high density, forming a non-singular black hole
\cite{ClosedHorHighDer,ClosedHor,ClosedHorTrapping,BHInt,Evap,Modesto,ModestoConn}
just as a cosmological model might bounce at the big bang. However,
homogeneous minisuperspace models are unable to reveal the structure of
space-time that could correspond to their solutions because the only
symmetries they allow are time reparameterizations. The interplay of spatial
and temporal transformations, locally expressed by Lorentz boosts, remains
undetermined. In particular, in minisuperspace models it is impossible to tell
whether modifications such as (\ref{Cmod}) are compatible with general
covariance.

An interesting suggestion to circumvent this problem has been made in
\cite{Transfig}. (As a new model of quantum-modified black holes, this
proposal suffers from serious drawbacks as pointed out in
\cite{DiracPoly,TransCommAs,ExtendedPoly,bvPoly,MassPoly,LoopISCO} based on
several independent arguments. Here, we will be interested in possible
statements about space-time structure that are insensitive to properties of
specific solutions.)  The authors point out that even the inhomogeneous
exterior of a Schwarzschild black hole, by virtue of being static, allows a
slicing by homogeneous hypersurfaces, but they are timelike and therefore do
not correspond directly to anisotropic cosmological evolution. Nevertheless,
the canonical formulation as well as quantization, or modification as in
(\ref{Cmod}), can be performed on a timelike slicing. It is therefore possible
to explore implications of modified dynamics in the exterior, possibly
connecting it with the modified interior through a horizon in order to arrive
at a complete black-hole model.

The authors of \cite{Transfig} worked out many details of the resulting
solutions. However, they did not endeavor to determine the corresponding
space-time structure, or rather to test whether there is any well-defined
space-time structure at all. They instead used the same implicit assumption as
in \cite{PlanckStar,Fireworks} and postulated that modifications in their
anisotropic minisuperspace model can always be translated into a line element
with modified coefficients. Therefore, they presupposed that their model is
generally covariant without providing a proof.

As shown in \cite{TransComm,Disfig}, the model of \cite{Transfig} not only
fails to be covariant, it can even be used to derive a no-go theorem for the
covariance of any modification of the form (\ref{Cmod}). By taking these
modifications to the exterior, the resulting equations of motion become
sensitive to space-time structure for the following general reason: If a
region of space-time permits a timelike homogeneous slicing, it also permits a
static spherically symmetric slicing. These two slicings are related by an
exchange of time and space coordinates in the models, which allows one to
derive modified spherically symmetric dynamics from any proposed modified
homogeneous dynamics. The anisotropic but (timelike) homogeneous slicing
implies a line element of the form
\begin{equation} \label{dsh}
 {\rm d}s_{\rm homogeneous}^2= J(n)^2{\rm d}n^2- a(n)^2{\rm d}t^2+ b(n)^2
 \left({\rm  
     d}\vartheta^2+ \sin^2\vartheta{\rm d}\varphi^2\right)
\end{equation}
where $n$ is a coordinate in a direction normal to timelike slices, while the
time coordinate $t$ is part of the timelike homogeneous slices. The
coefficient $J(n)$ is the lapse (or, rather, spatial ``jump'') function in a
spacelike direction. 

For any choice of $J$, $a$ and $b$, this line
element is locally equivalent to the static spherically symmetric line element
\begin{equation} \label{dss}
{\rm d}s_{\rm spherically\;symmetric}^2= -a(x)^2{\rm d}t^2+
  J(x)^2 {\rm d}x^2+ b(x)^2 \left({\rm
      d}\vartheta^2+\sin^2\vartheta{\rm d}\varphi^2\right)
\end{equation}
if we just rename $n$ as $x$ and use spacelike slices of constant $t$ instead
of timelike slices of constant $n$. If a solution of the (modified)
homogeneous model on timelike slices presents a covariant space-time, it must
therefore be equivalent to a static spherically symmetric solution of some
spherically symmetric model, based on the transformation of metric components
implied by equating ${\rm d}s_{\rm
  spherically\;symmetric}^2$ with the generic form of a spherically symmetric
line element,
\begin{equation} \label{ds}
 {\rm d}s^2= -N(t,x)^2{\rm d}t^2+ L(t,x)^2\left({\rm d}x+M(t,x){\rm
     d}t\right)^2 + S(t,x)^2 \left({\rm
     d}\vartheta^2+\sin^2\vartheta{\rm d}\varphi^2\right)
\end{equation}
with free functions $N$, $M$, $L$ and $S$ depending only on time $t$ and the
radial coordinate $x$.

Unlike homogeneous models, spherically symmetric models are strongly
restricted by general covariance \cite{Strobl,BHSigChange}: If they are local,
quadratic in momenta and without higher derivatives, they must, up to field
redefinitions, be of the form of $1+1$-dimensional dilaton gravity, with
action
\begin{equation} 
 S[g,\phi]= \frac{1}{16\pi G} \int{\rm d}t{\rm d}x
 \sqrt{-\det g} 
 \left(\phi R- \frac{1}{2}
   g^{ab}\frac{\partial\phi}{\partial x^a}
     \frac{\partial\phi}{\partial x^b}  - V(\phi)\right)\,,
\end{equation}
in which only a single function, the dilaton potential $V(\phi)$, can be
changed in order to adjust the equations to potential modifications. 
If the condition of quadratic dependence on momenta is dropped, the form of
local generally covariant actions with second-order field equations is still
quite restricted, given by generalized dilaton models \cite{NewDilaton} of the
form 
 \begin{eqnarray}\label{S}
 {\rm S}[g,\phi]&=& \frac{1}{16\pi G} \int {\rm d}t{\rm d}x \sqrt{-\det g}
 \bigl(\xi(\phi) R+ k(\phi,X)\nonumber\\
&&+ C(\phi,X) \nabla^a\phi\nabla^b\phi
   \nabla_a\nabla_b\phi\bigr)\,.
\end{eqnarray}
Instead of a single dilaton potential, there are now three free functions,
$\xi(\phi)$, $k(\phi,X)$ and $C(\phi,X)$, but crucially they can depend only
on the scalar field $\phi$ and the first-order derivative expression
\begin{equation}
 X=-\frac{1}{2} g^{ab} \nabla_a\phi\nabla_b\phi\,.
\end{equation}
According to \cite{DilatonHorndeski}, the actions (\ref{S}) present the most
general 2-dimensional local scalar-tensor theories with second-order field
equations, or a 2-dimensional version of Horndeski theories \cite{Horndeski}.

The canonical structure of modifications such as (\ref{Cmod}), without
independent momenta not seen in the classical theory, characterizes
loop-modified models as local ones without higher derivatives. The modified
timelike homogeneous dynamics they imply must therefore be consistent with
some choice of generalized dilaton model with static solutions, if they have a
chance of being covariant. Importantly, the free functions in (\ref{S}) can
depend only on one of the degrees of freedom, $\phi$, and its first-order
derivatives but not on the 2-dimensional metric $g_{ab}$. In a spherically
symmetric interpretation, $\phi$ is determined by the function $S(t,x)$ in
(\ref{ds}).

A detailed analysis using the canonical equations of dilaton gravity shows
that it is impossible to express loop modifications in generalized dilaton
form \cite{TransComm,Disfig}. As a brief argument, holonomy modifications of a
model with line element (\ref{dsh}) imply a Hamiltonian that is non-polynomial
in the momenta of $a$ and $b$, and therefore non-polynomial in normal
derivatives of these coefficients by $n$. In the spherically symmetric
interpretation, normal derivatives of $a$ and $b$ are translated into spatial
derivatives of the lapse function $N$ and $S$ by $x$ in
(\ref{ds}). Non-polynomial corrections in $\partial N/\partial x$ cannot be
expressed in terms of the free functions of a generalized dilaton model. Loop
modifications of the timelike homogeneous model therefore cannot be consistent
with slicing independence.

The modified timelike-homogeneous dynamics in a classical space-time therefore
cannot be interpreted geometrically in terms of a metric or a line
element. The assessment, given in \cite{RovelliTrans}, that the authors of
\cite{Transfig} ``have shown that loop quantum gravity --- tentative theory of
quantum gravity --- predicts that spacetime continues across the center of the
hole into a new region that has the geometry of the interior of a white hole,
and is located in the future of the black hole'' is therefore incorrect
because in this model there is, in fact, no space-time that could describe the
modified solutions even at low curvature, let alone continue space-time across
the center of the black hole. It is not true that the authors ``have shown
that a crucial ingredient of this scenario, the transition at the center,
follows from a genuine quantum gravity theory, namely loop theory'' or that
``loop gravity predicts that the interior of a black hole continues into a
white hole'' \cite{RovelliTrans}.

Modifications of loop quantum cosmology are inconsistent with slicing
independence, which is a consequence of classical general covariance. In the
next section we will discuss a way to evade the conclusion that models of loop
quantum gravity violate covariance, based on the possibility that classical
symmetries may be deformed --- that is, affected by quantum corrections ---
even if they are not violated. Such a theory, using the generalization (ii.b')
of the classical condition (ii.b) in Section~\ref{s:Basic}, would still be
consistent and free of anomalies, but it would not permit an interpretation as
a geometric theory with solutions based on Riemannian geometry. Before we
discuss the underlying models, it is useful to consider another proposal that
aimed (but ultimately failed) to evade strong restrictions from general
covariance on modifications suggested by loop quantum gravity.

\subsubsection{Spherically symmetric models}
\label{s:SphSymm}

Before we are ready to discuss deformed covariance and space-time structures,
we should comment on several attempts to derive black-hole models by evading
the covariance question. Some of them simply ignore general covariance by
fixing the space-time gauge before implementing quantization or modification,
and then working only in this one gauge fixing
\cite{LoopCollapse,LoopGauge1,LoopGauge2}. These attempts need not be
discussed in any detail because modifying the equations of a gauge theory
after fixing the gauge leads to questionable physics unless it can be shown
that a covariant quantization of this kind exists.  Without an explicit
demonstration of gauge invariance, none of the conditions (i), (ii.a) and
(ii.b) or (ii.b') of Section~\ref{s:Basic} are guaranteed to hold.

A more refined suggestion to implement modifications similar to (\ref{Cmod})
has been made in \cite{LoopSchwarz}, using spherically symmetric models,
classically described by space-times with line elements (\ref{ds}). (The
original constructions of \cite{LoopSchwarz} were made in triad variables, but
they are independent of this choice and hold equally in metric variables as
used here.) The isotropic Friedmann constraint is then replaced by the
functional Hamiltonian constraint
\begin{equation} \label{H}
 H[N]= \int N\left(-\frac{p_Lp_S}{S}+
   \frac{Lp_L^2}{2S^2}+ 
\frac{1}{2} \frac{(S')^2}{L} +\frac{SS''}{L}-
 \frac{SS'L'}{L^2} + \frac{1}{4} LSV(S)\right){\rm d}x
\end{equation}
where $p_S$ and $p_L$ are canonically conjugate to $S$ and $L$, respectively.
Here, for the sake of generality, we have included the dilaton potential
$V(S)$ of $1+1$-dimensional dilaton gravity, which equals $V(S)=-2/S$ in
spherically symmetric models reduced from $3+1$-dimensional general
relativity. In addition, we have the diffeomorphism constraint
\begin{equation} \label{Diff}
 D[M]= \int M \left(S'p_S-Lp_L'\right)\,,
\end{equation}
such that
\begin{equation} \label{HH}
 \{H[N_1],H[N_2]\}=D[L^{-2}(N_1N_2'-N_1'N_2)]\,.
\end{equation}
The last condition ensures that gauge transformations generated by the
Hamiltonian and diffeomorphism constraints correspond to hypersurface
deformations in some classical space-time with Riemannian geometry. In any
spherically symmetric theory with (\ref{HH}), condition (ii.b) of
Section~\ref{s:Basic} holds.

Modifying the Hamiltonian constraint of spherically symmetric models is a much
more non-trivial exercise than modifying the single constraint of isotropic
cosmological models. Most attempts simply break the underlying classical
symmetries, such that the Poisson bracket $\{H[N_1],H[N_2]\}$ is not related
to any of the relevant generators, $H[N]$ and $D[M]$. An interesting
observation was made in \cite{LoopSchwarz} which makes it easier to
incorporate modified constraints: The linear combination
\begin{equation} \label{HDP}
 H[2P S'/L]+D[2P p_L/(SL)]= \int P
 \frac{{\rm d}}{{\rm d}x} \left(-\frac{p_L^2}{S}+ \frac{S(S')^2}{L^2}+
   \frac{1}{2} \int SV(S){\rm 
     d}S\right){\rm d}x
\end{equation}
of constraints does not depend on $p_S$, and the remaining terms form a
complete spatial derivative, except for the new multiplier $P$. As a
constraint, the linear combination can therefore be replaced by a new
constraint,
\begin{equation} \label{CQ}
 C[Q]= \int Q \left( -\frac{p_L^2}{S}+
   \frac{S(S')^2}{L^2}+ 
   \frac{1}{2} \int SV(S){\rm 
     d}S+ C_0\right){\rm d}x\,,
\end{equation}
with a free constant $C_0$, and $Q$ is now a multiplier with density weigtht
minus one. 

The density weight implies that the bracket of $C[Q]$ with the
diffeomorphism constraint is
\begin{equation} \label{CD}
 \{C[Q],D[M]\}=-C[(M Q)']\,,
\end{equation}
while it is easy to see that
\begin{equation}  \label{CC}
 \{C[Q_1],C[Q_2]\}=0
\end{equation}
because $C[Q]$ depends only on the momentum $p_L$ and on none of the spatial
derivatives of $L$. The Poisson bracket therefore produces only delta
functions but none of their spatial derivatives, and all terms
cancel out in the antisymmetric $\{C[Q_1],C[Q_2]\}$.  A non-Abelian constraint
$H[N]$ with structure functions in the bracket (\ref{HH}) has therefore been
replaced by a partially Abelian constraint $C[Q]$ with structure constants.

The general properties of (\ref{CQ}) that imply (\ref{CC}) remain unchanged if
$p_L^2$ in (\ref{CQ}) is replaced with an arbitrary function of $p_L$. The
modified constraint then still depends only on the momentum $p_L$, and not on
spatial derivatives of $L$. Moreover, since $p_L$ has density weight zero, the
bracket (\ref{CD}) is not affected by the modification. It therefore seems
possible to implement an arbitrary $p_L$-dependent modification of (\ref{CQ})
without changing the brackets of constraints, just as it is possible to modify
(\ref{Cmod}) in isotropic models. In particular, the modified constraints
remain first class (they obey condition (ii.a) of Section~\ref{s:Basic}) and
could be expected to describe a generally covariant theory that could be
expressed in terms of modified line elements.

However, this conclusion ignores the non-trivial nature of
hypersurface-deformation brackets, or of condition (ii.b)
\cite{SphSymmCov}. For a covariant theory, it is not sufficient to have an
anomaly-free system of constraints with closed brackets such as (\ref{CD}) and
(\ref{CC}). The generators should also correspond to hypersurface deformations
in space-time. Since the partially Abelianized brackets (\ref{CD}) and
(\ref{CC}) are not of hypersurface-deformation form, for a covariant
space-time theory we must be able to form linear combinations of the
constraints such that a bracket of the form (\ref{HH}) is obtained, providing
the correct relationship required for normal deformations of hypersurfaces in
space-time \cite{DiracHamGR,ADM,Regained}.  Without this demonstration, the
model obeys condition (ii.a) but not condition (ii.b) from
Section~\ref{s:Basic}, and therefore could not be used to derive effective
line elements.

After modifying the Abelianized constraint $C$, as in
\begin{equation} \label{CQf}
 C[Q]= \int Q \left( -\frac{f_1(p_L)}{S}+
   \frac{S(S')^2}{L^2}+ 
   \frac{1}{2} \int SV(S){\rm 
     d}S+ C_0\right){\rm d}x
\end{equation}
with some modification function $f_1$, we must therefore be able to retrace
our steps that led to the definition of $C$ if the modification is to preserve
general covariance. Given the form of the diffeomorphism
constraint\footnote{The diffeomorphism constraint is determined by the
  condition that the integrated expression $C[Q]$, in order to be meaningful,
  must have an integrand with density weight one. This condition implies the
  density weights of all basic canonical fields, which in turn uniquely
  specify the diffeomorphism constraint up to adding a constant
  \cite{Foundations}. This implied diffeomorphism constraint is unmodified and
  equals the classical constraint. The fact that loop quantum gravity does not
  provide an operator for the diffeomorphism constraint, but only operators
  for finite diffeomorphisms, does not matter here because an effective model
  implicitly works with expectation values taken in a suitable (here,
  near-continuum) class of states. (Thanks to Jorge Pullin for bringing up the
  lack of a diffeomorphism constraint operator in private correspondence.)}
and the known expression of the classical Hamiltonian constraint, which should
be obtained in the ``classical'' limit where $f_1(p_L)\to p_L^2$, the only
combination that could equal the modified $C[Q]$, or the derivative expression
(\ref{HDP}) from which it is derived, is
\begin{eqnarray}
 &&  \int P
 \frac{{\rm d}}{{\rm d}x} \left(-\frac{f_1(p_L)}{S}+ \frac{S(S')^2}{L^2}+
   \frac{1}{2} \int SV(S){\rm 
     d}S\right){\rm d}x\nonumber\\
&=& \int 2P
 \left(- \frac{p_L'}{2S} \frac{{\rm d}f_1}{{\rm d}p_L}  
 +\frac{S'}{L}\left(\frac{Lf_1(p_L)}{2S^2}+
     \frac{(S')^2}{2L}+ \frac{SS''}{L}- \frac{SS'L'}{L^2}+ 
   \frac{LSV(S)}{4}\right)\right)  {\rm d}x\nonumber\\
&=& \tilde{H}[2P S'/L]+D[2P f_2(p_L)/(SL)] \label{HDC}
\end{eqnarray}
with some function $f_2(p_L)$ in the multiplier of the diffeomorphism
constraint, and a possibly modified Hamiltonian constraint $\bar{H}[N]$.  This
equation, together with an unmodified diffeomorphism constraint because the
canonical formulation assumes the classical structure of space at equal times,
implies that 
\begin{equation}\label{f1f21}
 f_2(p_L)= \frac{1}{2} \frac{{\rm d}f_1}{{\rm d}p_L}\,.
\end{equation} 
This function is uniquely determined by the single term in (\ref{HDC}) that
depends on $p_L'$. Subtracting $D[2P f_2(p_L)/(SL)]$ from (\ref{HDC}) then
implies
\begin{equation} \label{Hf}
 \tilde{H}[N]= \int  N\left(-\frac{f_2(p_L)p_S}{S}+
   \frac{Lf_1(p_L)}{2S^2}+ 
\frac{1}{2} \frac{(S')^2}{L} +\frac{SS''}{L}-
 \frac{SS'L'}{L^2} + \frac{1}{4} LSV(S)\right){\rm d}x\,.
\end{equation}

This modified constraint is free of anomalies for any function $f_1(p_L)$,
such that $f_2$ is determined by (\ref{f1f21}), because it has just been
derived as a linear combination of constraints in an anomaly-free, partially
Abelianized system. But the restrictive nature of general covariance in the
form of hypersurface-deformation generators, canonically encoding condition
(ii.b) of Section~\ref{s:Basic}, can nevertheless be seen.  For instance, if
$f_1$ depends on $S$ in addition to $p_L$, as suggested in
\cite{LoopSchwarzBar}, such as $S^{-2j}f_1(p_LS^j)$ with some constant $j$
which may be subject to renormalization, the combination used in (\ref{HDC})
includes a term $-\frac{1}{2} jS^{-j-2} S' p_L ({\rm d}f_1/{\rm
  d}z)|_{z=p_LS^j}$. Hypersurface-deformation generators can still be
reconstructed from a modified (\ref{CQf}) with a similar relationship,
$f_2(p_LS^j)= \frac{1}{2}S^{-j} ({\rm d}f_1/{\rm d}z)|_{z=p_LS^j}$, between
the two modification functions in (\ref{CQf}) and in the multiplier of the
diffeomorphism constraint, respectively.

However, the reconstructed modified Hamiltonian constraint is of the form
(\ref{Hf}) with $f_1$ replaced by
\begin{equation}\label{fff}
 \bar{f}_1(p_LS^j)= (2j+1) \frac{f_1(p_LS^j)}{S^{2j}}- 2j p_L f_2(p_LS^j) \,.
\end{equation}
While the resulting modified system is anomaly-free, it is not possible for
all three functions in (\ref{fff}), $f_1$, $f_2$ and $\bar{f}_1$, to be
periodic in their argument. It is therefore more difficult to motivate these
modifications by holonomy terms. This subtlety had already been pointed out in
\cite{SphSymmMaxwell,HigherSpatial}, but it went unnoticed in
\cite{LoopSchwarzBar} because general covariance was not analyzed in this
paper (while effective line elements were nevertheless proposed). In certain
other systems that may be partially Abelianized, such as spherically symmetric
gravity with a scalar field \cite{LoopSchwarz2}, one can show that no
hypersurface-deformation generators exist \cite{SphSymmCov}. Even though
anomaly-free quantizations or modifications can then be found in the partially
Abelian system, they cannot be considered covariant.

In the vacuum case, while the modified constraint (\ref{Hf}) is anomaly-free,
it obeys a bracket
\begin{equation} \label{HHbeta}
 \{\tilde{H}[N_1],\tilde{H}[N_2]\}= D[L^{-2} \beta(p_L) (N_1N_2'-N_1'N_2)]
\end{equation}
where
\begin{equation} 
 \beta(p_L)= \frac{1}{2} \frac{{\rm d}^2f_1}{{\rm d}p_L^2}
\end{equation}
if $f_1$ depends only on $p_L$ (and a related expression if $f_1$ depends on
both $p_L$ and $S$ through the combination $p_LS^j$; see \cite{JR}).  If
$\beta(p_L)\not=1$, the symmetries are modified compared with the classical
relationship (\ref{HH}). They no longer correspond to hypersurface
deformations in space-time, and are not dual to coordinate
transformations. The model of \cite{LoopSchwarz} is not consistent with
condition (ii.b) of Section~\ref{s:Basic}, but it is compatible with the
generalized form (ii.b'). Modifying partially Abelianized constraints
(\ref{CQ}) therefore produces modified space-time structures in disguise. It
is then not guaranteed that line elements obtained by inserting solutions of
modified constraints are meaningful because such solutions would be subject to
gauge transformations that are not dual to coordinate changes of ${\rm
  d}x^a$. A detailed analysis of modified space-time structures is therefore
required.

\section{Modified space-time structure}
\label{s:Mod}

Well before \cite{LoopSchwarz} was published, modified brackets of the form
(\ref{HHbeta}) had already been found by analyzing space-time structure
directly in terms of generators of hypersurface deformations
\cite{JR,LTBII,HigherSpatial,SphSymmOp}. These studies followed the usual
reasoning of effective field theories, in which potential quantum effects are
explored in a theory of classical type by including quantum modifications as
well as other terms of the same order that are consistent with all required
symmetries, here given by general covariance as represented by hypersurface
deformations. For a detailed discussion of effective field theory applied to
the canonical formulation of spherically symmetric models, see
\cite{EuclConn}.

\subsection{Anomaly-freedom}

In a Lagrangian treatment, an effective derivation of the covariance condition
usually leads to higher-curvature effective actions, perhaps with additional
independent degrees of freedom in terms of new fields. Symmetries that
determine the structure of allowed contributions to an effective action up to
a given order in derivatives can be evaluated systematically by using the
tensor-transformation law. Higher-dervative terms must then be curvature
invariants, or suitable combinations of derivatives of new fields.

In a Hamiltonian formulation, considerations of the {\em space-time} tensor
transformation law cannot be used directly because space-time fields are
decomposed into components according to a foliation of space-time. Tensor
transformations are replaced by closure conditions imposed on the Poisson
brackets of the Hamiltonian and diffeomorphism constraint, which directly
refer to the central objects of a Hamiltonian formulation. A canonical theory
is generally covariant provided it has constraints that generate hypersurface
deformations, such that (\ref{HH}) is fulfilled
\cite{Regained}. Higher-curvature effective actions indeed obey this
relationship, irrespective of their coefficients of higher-curvature terms
\cite{HigherCurvHam}.

The Hamiltonian treatment is technically more involved than the Lagrangian
one, but it is important because it also allows for generalizations of
covariance: The closure condition is more general than an application of the
tensor transformation law because the latter, but not the former, presupposes
that space-time is equipped with Riemannian geometry. In the language of
hypersurface deformations, Riemannian geometry is equivalent to the choice of
$\beta=1$ in (\ref{HHbeta}), or $\beta=-1$ in Euclidean signature. The
existence of consistent spherically symmetric models with $\beta\not=\pm 1$
proves that the Hamiltonian treatment is more general than Lagrangian
considerations based on the tensor-transformation law of Riemannian geometry.

When one tries to implement holonomy modifications in spherically symmetric
models using the language of effective field theory, one can start with the
expression (\ref{Hf}) already encountered in the preceding section. The two
functions $f_1$ and $f_2$ are initially arbitrary and allow for a modified,
non-quadratic dependence on the momentum $p_L$. While the dependence on $p_S$
could also be expected to be modified, no covariant version of this form has
been found yet. A possible explanation \cite{HigherSpatial} is that $p_S$,
unlike $p_L$, has a spatial density weight in spherically symmetric
space-times, and therefore needs to be integrated spatially before it can be
inserted in a non-linear function with well-defined spatial transformation
properties. Such modifications would therefore be non-local or, in a
derivative expansion, give rise to higher spatial derivatives. For such terms
to be consistent in an effective treatment, one would also have to include a
series of higher-derivative modifications of the spatial derivatives of $L$
and $S$ already present in (\ref{Hf}), considerably complicating derivations
of the Poisson brackets of constraints. It is therefore more difficult to find
modifications of the $p_S$-dependence, but such a modification has not been
ruled out yet. (Similar considerations apply to studies that aim to modify the
full theory because all components of the gravitational momentum then have
non-scalar transformation properties with respect to spatial coordinate
changes. Simple modifications that do not include higher spatial derivatives
have indeed been ruled out \cite{DefGenBH}.)

Two Hamiltonian constraints of the form (\ref{Hf}) have
a Poisson bracket 
\begin{equation} 
 \{\bar{H}[N_1],\bar{H}[N_2]\}= \int (N_1N_2'-N_1'N_2)
 \left(\frac{S'}{LS} \left(f_2(p_L)- \frac{1}{2} \frac{{\rm d}f_1}{{\rm
         d}p_L} \right) + \frac{1}{L^2} \frac{{\rm d}f_2}{{\rm d}p_L}
   \left(p_SS'-Lp_L'\right)\right){\rm d}x \,.
\end{equation}
The first term on the right-hand side does not vanish on the constraint
surface of Hamiltonian and diffeomorphism constraints. It must therefore
vanish for an anomaly-free implementation of modifications, which requires
\cite{JR}
\begin{equation} \label{f1f2}
 f_2(p_L)=\frac{1}{2} \frac{{\rm d}f_1}{{\rm d}p_L}\,.
\end{equation}
This is the same condition (\ref{f1f21}) found in the preceding section,
but derived independently.

The Poisson bracket of two Hamiltonian constraints is then
proportional to the diffeomorphism constraint, and there are no
anomalies. It is of the form (\ref{HHbeta}) with
\begin{equation} \label{beta}
 \beta(p_L)= \frac{{\rm d}f_2}{{\rm d}p_L}= \frac{1}{2} \frac{{\rm
     d}^2f_1}{{\rm d}p_L^2}\,, 
\end{equation}
a function that, in general, is not equal to $\pm 1$. The consistent
modifications in (\ref{Hf}) with (\ref{f1f2}) therefore imply a non-Riemannian
space-time structure. For small $p_L$, $\beta\to1$ provided $f_1(p_L)\to
p_L^2$ to respect the classical limit. If $f_1$ is a bounded function with a
local maximum, we have $\beta<0$ around the local maximum, which corresponds
to a non-classical version of space-time with Euclidean signature. For
instance, if $f_1(p_L)=\delta^{-2} \sin^2(\delta p_L)$ as in the isotropic
$C_{\rm modified}$, we have
\begin{equation}
 \beta(p_L)= \cos(2\delta p_L) \approx -1
\end{equation}
around local maxima of $f_1$, where $\delta p_L= (k+1/2)\pi$ with integer $k$.

The modified Hamiltonian constraint (\ref{Hf}) together with the usual
diffeomorphism constraint (\ref{Diff}) define a canonical system that is free
of anomalies and covariant in a generalized sense \cite{Action}, respecting
condition (ii.b') of Section~\ref{s:Basic}: The bracket (\ref{HHbeta}),
together with the unmodified classical bracket between $\bar{H}[N]$ and
$D[M]$, shows that the set of constraints is not only closed algebraically,
but also such that hypersurface deformations are obtained in the classical
limit, $\beta\to 1$. The second property defines generalized covariance, and
it is not implied by the first property, the closure condition.

\subsection{Signature change}

For $\beta\not=1$ in (\ref{HHbeta}), algebraic relations between the
constraints are modified compared with the classical form of hypersurface
deformations. Therefore, the transformations they generate cannot be
interpreted as changes of tensor fields on a Riemannian space-time canonically
foliated by hypersurfaces. In this way, generalized covariance is able to
evade the no-go results discussed in Section~\ref{s:Slicing} because the
latter assume the classical form of general covariance which is equivalent to
slicing independence in space-time. At present, it is not known how to
interpret the transformations of generalized covariance in geometrical
language under all circumstances, but for solutions such that $\beta$, through
$p_L$, depends only on time it has been shown in \cite{EffLine} that a field
redefinition of metric components can be used to map the space-time
description to Riemannian form. This case may be used in the interior region
of a Schwarzschild black hole within the horizon, using a gauge that implies
homogeneous spatial slices. It is therefore instructive in the present
context, even though the inhomogeneous exterior where $\beta$ depends on the
spatial position appears to require a more involved geometrical description in
which, according to the no-go results of Section~\ref{s:Slicing}, slicing
independence cannot easily be seen.

In the interior, the field redefinition derived in \cite{EffLine} leads to an
effective line element that is compatible with modified gauge transformations
generated by (\ref{Hf}). The space-time line element requires transformations
of all metric components, not just of the spatial part $q_{ab}$ which provides
phase-space degrees of freedom and can directly be transformed by computing
the Poisson bracket
\begin{equation}  \label{Gaugeq}
  \{q_{ab},H[\epsilon^0]+D[\epsilon^i]\}= {\cal L}_{\xi} q_{ab}
\end{equation}
where $\epsilon^0$ and $\epsilon^i$ are such that the space-time vector field
$\xi^a$ that appears in the Lie derivative ${\cal L}_{\xi}$ has components
\begin{equation} 
 \xi^0 = \frac{\epsilon^0}{N} \quad,\quad \xi^i = \epsilon^i-\frac{N^i}{N}
 \epsilon^0\,.
\end{equation}
(These components rewrite the vector field $(\epsilon^0,\epsilon^i)$ expressed
in a basis adapted to the foliation by using the unit normal $n^a$ into
components with respect to a coordinate basis with the usual ADM-like time
direction $t^a=Nn^a+N^a$ \cite{ADM} with the lapse function $N$ and spatial
shift vector $N^a$ on a given space-time on which the transformation is
performed \cite{PhaseSpaceCoord}.)

The remaining components of the space-time metric, given by lapse and shift in
\begin{equation}
 {\rm d}s^2= -N^2{\rm d}t^2+ q_{ij}\left({\rm d}x^i+N^i{\rm d}t\right)
 \left({\rm d}x^j+N^j{\rm d}t\right) \,,
\end{equation}
are subject to gauge transformations derived from the condition that
(\ref{Gaugeq}) is consistent with the time direction determined by these
functions. Based on this condition, lapse and shift transform according to
\cite{CUP}
\begin{equation}
 \delta_{\epsilon}N^a(x)= \dot{\epsilon}^a(x)+ \int{\rm d}^3 y{\rm d}^3z
 N^b(y)\epsilon^c(z) F_{bc}^a(x,y,z) 
\end{equation}
where $F_{bc}^a(x,y,z)$ are (distributional) structure functions of the
constraints, $C_a[\Lambda^a]=H[\Lambda^0]+D[\Lambda^i]$, such that
\begin{equation}
 \{C_a[\Lambda_1^a],C_b[\Lambda_2^b]\}= C_c[\Lambda_{12}^c]
\end{equation}
with
\begin{equation}
 \Lambda_{12}^c(x)= \int{\rm d}^3y{\rm d}^3z F_{ab}^c(x,y,z) \Lambda_1^a(y)
 \Lambda_2^c(z)\,.
\end{equation}

A modified bracket (\ref{HHbeta}) implies a non-classical $F_{00}^i$, and
therefore changes the gauge transformation of the shift components, in
addition to modified gauge transformations implied by (\ref{Gaugeq}) with a
modified (\ref{Hf}).  As shown by detailed derivations in \cite{EffLine} for
the case of a spatially constant $\beta$ in spherical symmetric space-times,
the new gauge transformations are dual to coordinate changes of ${\rm d}x^a$,
such that a suitable effective line element ${\rm d}s^2=\bar{g}_{ab}{\rm
  d}x^a{\rm d}x^b$ is invariant. The components of the effective metric
$\bar{g}_{ab}$ are of the form (\ref{ds}), except that $N^2$ is replaced with
$\beta N^2$ \cite{EffLine}:
\begin{equation} \label{dsbeta}
 {\rm d}s^2= -\beta(t) N(t,x)^2{\rm d}t^2+ L(t,x)^2\left({\rm d}x+M(t,x){\rm
     d}t\right)^2 + S(t,x)^2 \left({\rm
     d}\vartheta^2+\sin^2\vartheta{\rm d}\varphi^2\right)\,.
\end{equation}
An effective line element therefore exists in this case which expresses
generalized covariance in terms of Riemannian geometry after a field
redefinition. 

This line element shows directly that $\beta<0$ implies a transition to
Euclidean signature. The line element is degenerate when $\beta=0$ and
therefore does not describe a valid Riemannian geometry at the transition
hypersurface between Lorentzian and Euclidean signature. It therefore provides
two distinct effective descriptions of a single canonical solution,
corresponding to the regions where $\beta>0$ and $\beta<0$,
respectively. These two regions can be bridged in the canonical theory, whose
equations remain valid at $\beta=0$ \cite{EffLine,DefSchwarzschild,DefGenBH},
but there is no Riemannian interpretation of the transition surface. Since the
two regions are separated by a hypersurface of codimension one, it is possible
to extend fields across the transition surface by taking limits in one region
approaching the surface, and using the limiting values as initial conditions
for an extension into the other region. However, the existence of such a
mathematical extension does not necessarily imply a causal relationship.

\subsection{Signature change and non-singular space-time}

\begin{figure}
  \begin{center}
\includegraphics[width=10cm]{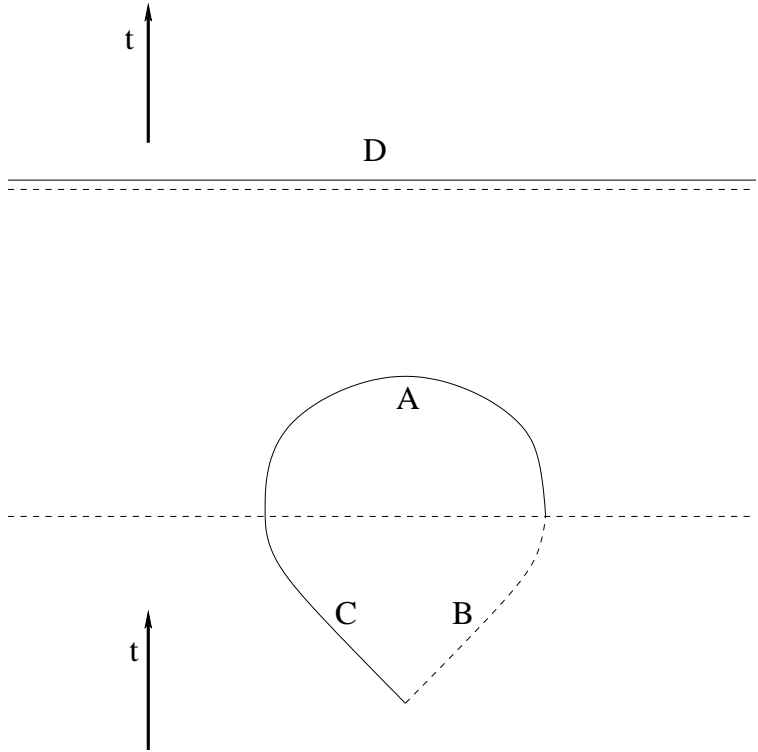}
\caption{Well-posed initial-boundary value problem for mixed-type partial
  differential equations in two dimensions, according to Tricomi
  \cite{Tricomi}. Arrows indicate the assumed flow of time in Lorentzian
  regions, while the two horizontal dashed lines are transition surfaces
  between two Lorentzian regions (top and bottom) and a Euclidean central
  region, as it may be realized at the center of a bouncing cosmological or
  black-hole solution. Entering the Euclidean region (bottom), the Tricomi
  problem shows that smooth data for the field (but not its normal
  derivatives) posed on the union of one characteristic in the original
  Lorentzian region (C) as well as an arc (A) in the Euclidean region
  connecting the end point of C with the end point of the other characteristic
  (B) starting at the same point as C imply a unique smooth solution in the
  region bounded by A, B and C, which depends continuously on the data. (The
  solution cannot be freely specified on B.) Data on A amount to final values
  as seen from the flow of time in the initial Lorentzian region. On the other
  side of the Euclidean region, top, the flow of time points away from the
  transition surface. A standard initial-value problem can therefore be used
  on any equal-time hypersurface (D) after the transition (``after'' defined
  according to the flow of time in the Lorentzian region pointing away from
  the Euclidean region). In this initial-value problem, both the field and its
  normal (that is, time) derivative can be chosen freely on
  D.  \label{f:Tricomi}}
\end{center}
\end{figure}

Fields on a space-time with line element (\ref{dsbeta}) obey partial
differential equations of mixed type, which are hyperbolic in the Lorentzian
region and elliptic in the Euclidean region \cite{SigImpl}. A well-posed
problem for solutions therefore requires a mixture of initial values in the
Lorentzian region and boundary values in the Euclidean region, which latter
appear as final conditions in a temporal interpretation based on the
Lorentzian phase; see Fig.~\ref{f:Tricomi}. The transition surface therefore
cannot be bridged by deterministic evolution, even though continuous and
well-behaved mathematical extensions are possible for given final
conditions. The precise form of relevant initial-boundary value problems has
been specified by Tricomi \cite{Tricomi}.

It has occasionally been claimed that signature change in models of loop
quantum cosmology has been ruled out by observations, but such arguments are
based on a single discredited study, \cite{QCExclude}, which erroneously used
a standard initial-value problem throughout the high-density phase. It is not
necessary to compute a power spectrum, as done in \cite{QCExclude}, in order
to rule out an ill-posed problem because of implied
instabilities. Discrepancies in the power spectrum derived in \cite{QCExclude}
compared with observations therefore do not rule out signature change; they
are only a symptom of the incorrect treatment of initial values. A
clarification of this problem has been published by some of the authors in
\cite{QCTime}. (The relevant issues had already been described in
\cite{SigImpl}.) In this context, it is important to note that signature
change is ubiquitous in models of loop quantum gravity. It appears not only in
a direct treatment of modified constraints of the form (\ref{Hf}), but is also
realized in a hidden way in partially Abelianized treatments based on
(\ref{CQf}), as described in Section~\ref{s:SphSymm}.

Signature change reveals a new possibility to avoid singularity theorems of
general relativity. Unlike a simple loop-motivated bounce based on
(\ref{Cmod}), it clearly shows which of the usual assumptions of these
theorems no longer hold. The main theorems are insensitive to the actual
dynamics and only use energy conditions as well as properties of Riemannian
geometry such as the geodesic deviation equation, in addition to topological
assumptions. It is then difficult to understand how vacuum solutions for black
holes can be singularity free, as in bouncing proposals, based on modified
dynamics while maintaining Riemannian geometry and (because there is no
matter) energy conditions. Of course, it is always possible to write
corrections to Einstein's equation as an effective stress-energy tensor, which
may formally violate energy conditions, but such a reformulation is not
necessary and therefore does not show how singularity theorems are
evaded. Moreover, rewriting a modification of Einstein's equation through an
effective stress-energy tensor assumes that the theory is covariant and
compatible with Riemannian geometry, which is not guaranteed.

\begin{figure}
\begin{center}
\includegraphics[width=8cm]{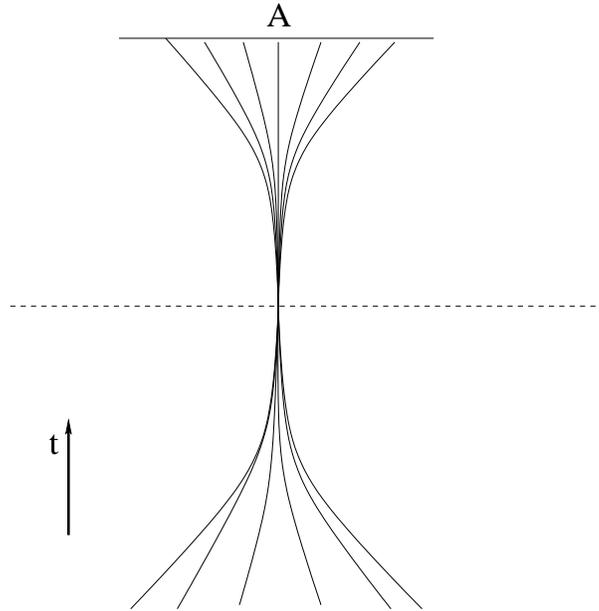}
\caption{Final conditions required for an extension of geodesics through a
  hypersurface of signature change (dashed). A $\beta$ approaching zero in
  the effective line element (\ref{dsbeta}) implies that light cones in the
  Lorentzian region collapse. All timelike geodesics aimed toward a given point
  on the transition surface therefore arrive there with the same asymptotic
  direction. Using the limiting values as initial values for extended (now
  spacelike) geodesics in the Euclidean region, a unique extension follows
  only if additional data are provided, such as the final point of a spacelike
  geodesic. The final points for a family of geodesics, lined up in this
  figure along a curve A, correspond to the final data on the arc A in the
  Tricomi problem, Fig.~\ref{f:Tricomi}.
  \label{f:Geodesics}}
\end{center}
\end{figure}

While scenarios of dynamical signature change are non-singular, unlike
classical signature change
\cite{ClassSigChange,SigChangeJunction,BoundarySigChange,KleinSigChange,ClassSigChange2},
they do not violate singularity theorems. Modified hypersurface deformations
according to (\ref{HHbeta}) imply that the entire space-time structure,
including both Lorentzian and Euclidean regions, is not Riemannian.  For this
reason, the usual theorems do not apply. Riemannian geometry can be used after
a field redefinition that leads to the effective line element (\ref{dsbeta}),
but only in two disjoint regions of Lorentzian and Euclidean signature,
respectively. If we start in the effective Lorentzian region, timelike
geodesics are indeed inextendible, as required by singularity theorems,
because the transition surface of signature change is reached after a finite
amount of proper time. Thereafter, time, and therefore timelike geodesics, do
not exist. However, the transition surface is not a boundary of space(-time)
in the generalized sense that allows for signature change. Geodesics can be
extended across this surface as spacelike ones, as illustrated in
Fig.~\ref{f:Geodesics}. Such an extension requires final values in the
Euclidean region, just as shown by the Tricomi problem for well-posed problems
of mixed-type partial differential equations.

\subsection{Evaporation scenarios ruled out by signature change}

\begin{figure}
\begin{center}
\includegraphics[width=4cm]{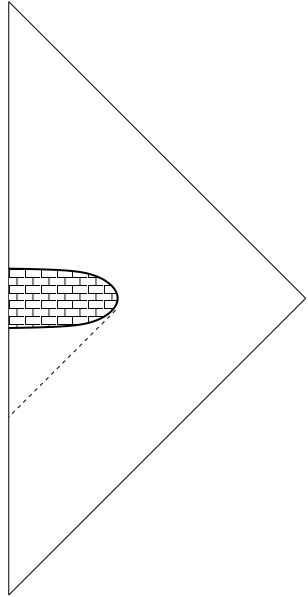}
\caption{A bouncing interior of a Planck star is ruled out by signature change
  because the high-density region (indicated as a brick wall) is Euclidean and
  does not allow deterministic evolution. Boundary values around the Euclidean
  region, required for a well-posed problem of Tricomi-type, imply
  indeterministic behavior as shown in Fig.~\ref{f:Cauchy}.
  \label{f:Bounce}}
\end{center}
\end{figure}

Based on loop quantum gravity, bounce-based black holes as suggested in
\cite{PlanckStar} are ruled out by signature change. While there may be no
singularity in a space-time modified by effects from loop quantum gravity, the
structure of space-time is modified such that deterministic evolution through
high curvature is impossible. The putative bounce as a dynamical process is
stopped in its tracks by signature change and does not happen. Matter that
collapsed or fell into the black hole does not reappear later because it
cannot evolve through timeless high density or curvature. While a white hole
might open up in the future where the black hole had been, the necessity of
final conditions, according to Figs.~\ref{f:Tricomi} or \ref{f:Geodesics},
implies that it is not uniquely determined by the previous state through the
complete lockdown imposed by the Euclidean phase. It therefore represents a
naked singularity: Specifying evolution to the future of the Euclidean region
requires a choice of new initial values on the top line (D) in
Fig.~\ref{f:Tricomi} which are not determined by the state of the initial
black hole; see Fig.~\ref{f:Bounce} for a possible embedding of the Euclidean
region in a space-time diagram.

\begin{figure}
\begin{center}
\includegraphics[width=4cm]{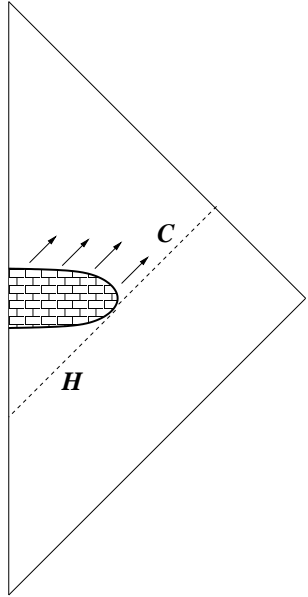}
\caption{A Euclidean core implies a Cauchy horizon $C$ in addition to the
  event horizon $H$, if it opens up in the
  original exterior. Observers crossing $C$ at low curvature will be exposed
  to undetermined information from the future boundary of the Euclidean region
  where new initial values must be posed for the future half of this Penrose
  diagram.  
  \label{f:Cauchy}}
\end{center}
\end{figure}

More generally, any black-hole interior reopening after the Euclidean lockdown
and connecting with the original exterior is ruled out because it would
violate deterministic behavior at low curvature: An exterior observer who
always stays at low curvature would suddenly be inundated by whatever data may
have been posed on the future boundary of the Euclidean region, unbeknownst to
the low-curvature observer. The rightmost edge of the Euclidean region is the
starting point of a Cauchy horizon \cite{Loss} because it marks the transition
into a region no longer determined by the distant past; see
Fig.~\ref{f:Cauchy}. Even though curvature remains finite, the top boundary of
the Euclidean region presents a naked singularity.

\subsection{Evaporation scenarios consistent with signature change}

Signature change does not always imply unacceptable violations of
deterministic behavior. There is certainly no determinism in a Euclidean
region because there is no time, but as long as this region is confined to
high curvature and does not have implications on observers who always stay at
low curvature, this behavior is not ruled out by common requirements on
fundamental physics.

\begin{figure}
\begin{center}
\includegraphics[width=4cm]{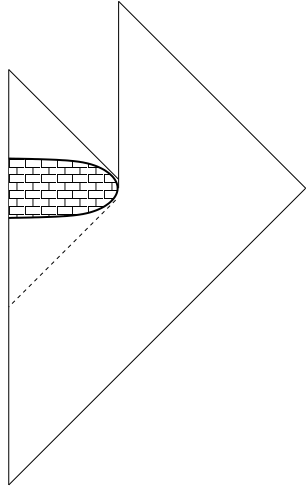}
\caption{An orphan universe starting at the future edge of
  the Euclidean core without connecting to the former black-hole exterior.
  \label{f:Baby}}
\end{center}
\end{figure}

A possible consistent scenario is given by a model in which the initial
black-hole interior does not open back up into the original exterior after the
Euclidean region. Instead, it forms an instantly orphaned baby universe that
lacks complete causal contact with a parent; see Fig.~\ref{f:Baby}. While data
on the future edge of the Euclidean region remain undetermined by the
black-hole past and technically play the role of a naked singularity, this is
no different from the initial singularity we are used to from classical
cosmological models. The naked singularity sets the stage for a new universe,
but, unlike in Fig.~\ref{f:Cauchy}, it does not affect observers who always
stayed at low curvature. This scenario is therefore permissible.

\begin{figure}
\begin{center}
\includegraphics[width=4cm]{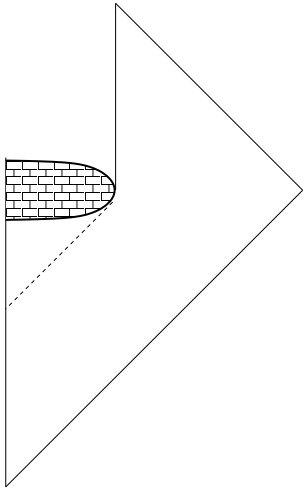}
\caption{The Euclidean region as a final boundary on which future data are
  posed and not evolved further. 
  \label{f:Final}}
\end{center}
\end{figure}

It is not necessary to include the orphan universe in the space-time
diagram. Instead, the Euclidean region could be a final boundary of the
black-hole interior on which future data are posed but not evolved further;
see Fig.~\ref{f:Final}. In this form, no naked singularity appears, or no
singularity at all because the Euclidean region which eliminates the usual
curvature singularity of Schwarzschild black holes now lacks a future
Lorentzian region and does not initiate a new space-time region unrelated to
its past. This consistent scenario is closely related to the independent
proposal of \cite{FinalState}, also explored in a related way in
\cite{Annihilation}.

While the scenarios of an orphan universe and a final state are consistent
with commonly accepted deterministic behavior, they do not solve the
information loss problem of black holes.

\subsection{Unexpected relationships with other approaches}

The final-state scenario already shows that models that take signature change
seriously lead to unexpected similarities between what has been suggested in
loop quantum gravity and other approaches. The absencce of bounces in
cosmology or in black holes is also surprisingly consistent with the
no-transition principle extracted in \cite{NTP} from a gauge-gravity
correspondence, which does not permit exchanges between independent quantum
field theories on a boundary. Although signature change is independent of
boundary components, it prevents the same kind of transitions as discussed in
\cite{NTP}. In \cite{Transfig}, by contrast, a contradiction between bouncing
black holes and the no-transition principle has been observed, claiming that
this is possible because in loop models ``one works {\em directly} in the
bulk'' (emphasis in \cite{Transfig}). However, a transition is possible in the
model of \cite{Transfig} only because this paper failed to capture the correct
bulk geometry, or rather to implement any consistent bulk geometry as per the
covariance conditions given in Section~\ref{s:Basic}.

The possibility of signature change (or mixed-type problems) in models of loop
quantum gravity is similar to certain higher-curvature theories such as
Einstein-dilaton Gauss--Bonnet gravity \cite{GBMixed,GBMixed2}; see also
\cite{HorndeskiWellPosed,HorndeskiNonHyp,HorndeskiHyp,HorndeskiWellPosedHarmonic}
for a general analysis in Horndeski theories \cite{Horndeski}. In models of
loop quantum gravity, signature change does not require higher-curvature terms
and is a result of modified space-time structures, not just of modified
dynamics. Nevertheless, there are interesting relationships between
implications for black holes if signature change can always be confined to
regions hidden behind horizons.

In the context of quantum gravity, different versions of signature change have
been found in a variety of approaches, including minisuperspace models
\cite{SigChangeMini}, string theory \cite{SigSingString}, matrix models
\cite{SigChangeFuzzy,SigChangeFuzzy2,SigChangeMatrix,SigChangeMatrixSol}, and
causal dynamical triangulations \cite{CDTSigchange}.

\subsection{Avoiding signature change in models of loop quantum gravity}

It is sometimes possible to construct holonomy modifications of certain models
without implying signature change. For instance, the Hamiltonian constraint
expressed using complex connections instead of the momenta in (\ref{H}) has a
different appearance of spatial derivatives of metric components. It can be
consistently modified in ways different from (\ref{Hf}), in particular such
that even the modified constraints obey (\ref{HHbeta}) with $\beta=1$
\cite{CosmoComplex,SphSymmComplex,GowdyComplex}. However, these modifications
are not generic because a complete effective description to the same order of
derivatives would include spatial derivative terms as in (\ref{H}), and
signature-changing brackets would result \cite{LQCScalarEucl,EuclConn}.
Working with generic modifications, the required relation (\ref{beta}) implies
signature change for any function $f_1$ that has a local maximum. Therefore,
signature change cannot be removed by familiar regularization options in loop
quantum gravity, in particular by choosing an ${\rm SU}(2)$-representation to
represent holonomies. Any such representation leads to a bounded and periodic
function $f_1$ which therefore has a local maximum. (Representation choices in
the context of deformed hypersurface-deformation brackets have been considered
in \cite{DeformSpinComplex,DeformSpin}.)

Another possibility to avoid signature change can be found if one allows for
non-bouncing solutions in loop quantum cosmology \cite{NonBouncing}, at least
in the context of cosmological perturbations. The bracket (\ref{HHbeta}) with
$\beta$ evaluated on a non-bouncing background solution is then still
modified, but $\beta>0$ throughout the high-density phase. A single Riemannian
geometry described by an effective line element (\ref{dsbeta}) then exists for
the entire transition. Since these solutions are non-bouncing, they reach zero
volume and may trigger a curvature singularity. However, the dynamics is
modified compared with classical cosmological models and could remain
non-singular, but this possibility has not been explored yet for the solutions
described in \cite{NonBouncing}. As supporting evidence for non-singular
behavior, it is known that the equations of loop quantum cosmology can remain
non-singular even in a transition through zero volume \cite{Sing}. What is at
present unclear is whether such a transition can be described in an effective
model suitable for the geometry of a black hole. 

Signature change --- or, more generally, a deformation of the
hypersurface-deformation brackets --- is derived in models of loop quantum
gravity from one specific quantum effect, given by holonomy modifications in
the Hamiltonian constraint. In addition, in a perturbative treatment of any
interacting theory one would expect a large number of quantum corrections from
loop diagrams which, in the case of gravity, usually come in the form of
higher-curvature corrections \cite{EffectiveGR,BurgessLivRev}. In a canonical
treatment, the corresponding higher-derivative terms can be seen to be implied
by quantum back-reaction of fluctuations and higher moments on the expectation
values of basic variables \cite{EffAc,Karpacz,HigherTime}. Based on general
algebraic properties of moments, derived from the quantum commutator, it can
be shown in general terms \cite{EffConsQBR} that quantum back-reaction on its
own cannot produce modified structure functions as in (\ref{HHbeta}), and
therefore does not interfere with any such modification implied by other
effects, such as holonomies. This result, which is consistent with the fact
that no higher-curvature effective action of general relativity modifies the
hypersurface-deformation brackets \cite{HigherCurvHam}, shows that signature
change is robust under the inclusion of quantum back-reaction or corrections
from loop diagrams. It is at present unknown how non-perturbative effects
might affect signature change or, more basically, the various effective
formulations currently used in all models of loop quantum gravity. The issue
of signature change (and how it could be avoided) provides strong motivation
for loop quantum gravity to focus on studying the challenging question of how
to understand non-perturbative, background-independent quantum effects in a
space-time picture.

Even if signature change could be avoided, its possibility is of conceptual
importance because it shows that proposed models that do not address (or even
explicitly violate) the covariance conditions spelled out in
Section~\ref{s:Basic} cannot be considered ``first approximations'' to some
complicated full theory of quantum gravity, as sometimes suggested. The
possibility of signature change shows that a failure to address the covariance
problem can have drastic consequences because it leads one to misinterpret the
causal structure consistent with one's proposed equations. By considering the
covariance problem, one includes strong consistency conditions that can rule
out certain proposals, as shown here.

\section{Conclusions}

A majority of previously proposed scenarios for evaporating black holes in
models of loop quantum gravity suffer from several severe problems. In
particular, they use equations that can be shown explicitly to violate general
covariance, and they are based on several crucial assumptions which have never
been demonstrated in this setting. An example of the latter is the postulate
that the interior of a black hole may bounce and open up again such that it is
causally connected to the former exterior. While bounces can be motivated by
more tractable cosmological models, using the well-known homogeneous slicing
of the Schwarzschild interior, showing that the interior reconnects to an
exterior space-time requires inhomogeneous models. In inhomogeneous models,
however, the equations of loop quantum gravity are much more complicated and
cannot be solved yet in any controlled way. If models are used that insert
potential loop effects into classical equations, there is a large number of
ambiguities as well as problems with covariance.

An interesting suggestion made in \cite{Transfig} initially indicated that
tractable homogeneous models could be applied even in the inhomogeneous
exterior, provided they are based on timelike slicings applied to static
solutions. However, loop modifications in the timelike slicing cannot be
compatible with general covariance \cite{TransComm,Disfig}. Instead of
producing a viable model of quantum black holes, the proposal of
\cite{Transfig} provided a crucial step in a demonstration that models of loop
quantum gravity violate general covariance.

Nevertheless, models of loop quantum gravity may be consistent provided they
incorporate a generalized version of covariance in which classical slicing
independence is replaced by a new quantum symmetry. In certain cases, field
redefinitions can be used to map the variables of such a model to an effective
metric which can be represented in the standard form of Riemannian geometry
and is consistent with slicing independence \cite{EffLine}. However, it
remains unknown whether such transformations are always possible. The precise
geometrical nature of generalized covariance is therefore unclear,
complicating detailed analyses of black-hole models in loop quantum gravity.

One common implication of generalized covariance in models of loop quantum
gravity is dynamical signature change, which usually takes place at large
curvature. Even while a precise geometrical description remains unknown,
corresponding consequences can be analyzed in general terms because signature
change implies a characteristic form of well-posed initial-boundary value
problems. When Euclidean regions are located to the future of a Lorentzian
low-curvature region, the boundary data they require for well-posed problems
amount to future data from the perspective of the Lorentzian region. They
imply the danger of violating deterministic behavior even for low-curvature
observers who never directly experience strong quantum-gravity effects. This
implication is sufficient to rule out the idea of Planck stars in models of
loop quantum gravity, given by bouncing black-hole interiors that reconnect
with the former exterior and may become visible. Two consistent options for
deterministic scenarios are given by an orphan universe in which a baby
universe splits up in a causally disconnected way, and a final-state scenario
in which the interior is not extended but the classical singularity is
replaced by a final condition.

\begin{figure}
\begin{center}
\includegraphics[width=6cm]{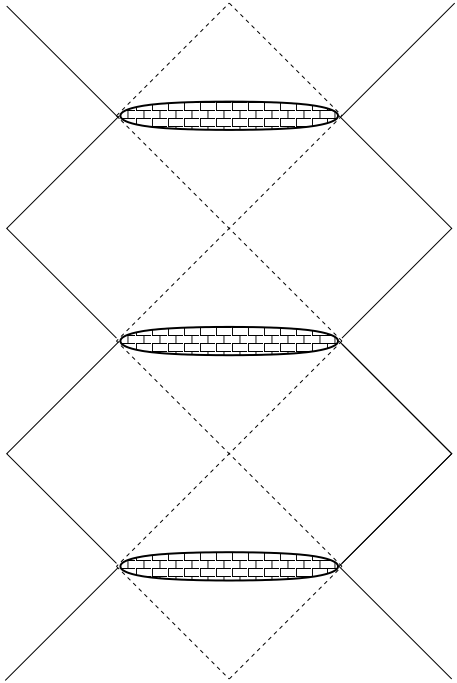}
\caption{Quantum extension of a Kruskal black hole. While curvature
  singularities may be removed in models of loop quantum gravity, they are
  replaced by regions of Euclidean space through which
  evolution is impossible. The extended space(-time) has unique solutions
  provided well-posed initial-boundary value problems are used for mixed-type
  equations. Lorentzian regions have finite boundaries at their transition
  hypersurfaces to Euclidean regions, which are non-singular but require
  boundary values that are interpreted as future data from the Lorentzian
  perspective. The extension shown here is derived from covariance together
  with consistency with
  deterministic behavior at low   curvature. 
  \label{f:Kruskal}}
\end{center}
\end{figure}

These scenarios of evaporating black holes have been described in
Section~\ref{s:Mod}. A consistent scenario for a vacuum black hole, extending
the classical Kruskal space-time, is shown in Fig.~\ref{f:Kruskal}, based on
the following two statements: (1) Interiors are non-singular, but they do not
bounce because evolution is blocked by Euclidean regions. (2) Future interiors
cannot reconnect with past exteriors because the boundary data required in
Euclidean regions would then lead to violations of deterministic behavior at
low curvature. The scenario shown in Fig.~\ref{f:Kruskal} is parsimonious: It
is consistent with low-curvature determinism, and it does not require a
complicated dynamical analysis to see whether solutions of loop quantum
gravity make it possible for a bouncing interior to reconnect with a former
exterior. The space(-time) required for such a scenario can instead be patched
together from non-singular interiors and almost-classical low-curvature
exteriors, with a resulting global structure that follows from general
consistency requirements. Even though global properties are derived, they are
implied by a careful local analysis of covariance in field equations.

In this scenario, future interiors are unobservable in past exteriors. The
model is therefore not directly testable by observations, but it is preferred
by a combination of internal consistency relations, including covariance and
low-curvature determinism. Based on these conditions, the scenario shown in
Fig.~\ref{f:Kruskal} presents the limit of what can at present be claimed in
models of loop quantum gravity. Future studies will have to show whether its
limitations to evolution implied by signature change are a consequence of
currently available effective descriptions, required for any detailed
analysis, or whether they hold even in a full theory of loop quantum gravity.

\section*{Acknowledgements}

This work was supported in part by NSF grant PHY-1912168.

%\bibliographystyle{../preprint}
%\bibliography{../Bib/QuantGra}

\end{document}